REVIEW **Open Access**

# Genes predisposing to syndromic and nonsyndromic infertility: a narrative review

Tajudeen O. Yahaya[1*], Usman U. Liman[2], Haliru Abdullahi[1], Yahuza S. Koko[1], Samuel S. Ribah[1], Zulkarnain Adamu[1] and Suleiman Abubakar[1]

## Abstract

**Background:** Advanced biological techniques have helped produce more insightful findings on the genetic etiology of infertility that may lead to better management of the condition. This review provides an update on genes predisposing to syndromic and nonsyndromic infertility.

**Main body:** The review identified 65 genes linked with infertility and infertility-related disorders. These genes regulate fertility. However, mutational loss of the functions of the genes predisposes to infertility. Twenty-three (23) genes representing 35% were linked with syndromic infertility, while 42 genes (65%) cause nonsyndromic infertility. Of the 42 nonsyndromic genes, 26 predispose to spermatogenic failure and sperm morphological abnormalities, 11 cause ovarian failures, and 5 cause sex reversal and puberty delay. Overall, 31 genes (48%) predispose to male infertility, 15 genes (23%) cause female infertility, and 19 genes (29%) predispose to both. The common feature of male infertility was spermatogenic failure and sperm morphology abnormalities, while ovarian failure has been the most frequently reported among infertile females. The mechanisms leading to these pathologies are gene-specific, which, if targeted in the affected, may lead to improved treatment.

**Conclusions:** Mutational loss of the functions of some genes involved in the development and maintenance of fertility may predispose to syndromic or nonsyndromic infertility via gene-specific mechanisms. A treatment procedure that targets the affected gene(s) in individuals expressing infertility may lead to improved treatment.

## Background

Infertility is generally defined as the inability of an organism to reproduce naturally. In humans, it is complex and defined as the failure to conceive after a year of regular and unprotected sexual intercourse [1]. Infertility affects about 48.5 million couples, representing 15% of couples worldwide [2]. Males are responsible for 20–30% of infertility, while females account for 20–35%, and the remaining is shared by both [2, 3]. However, the prevalence of infertility varies worldwide, being highest in South East Asia and West Africa [4, 5].

Infertility causes psychological, economic, and health burdens, resulting in trauma and stress, particularly in societies that emphasize childbearing [6]. In some parts of the world such as Africa and Asia, infertile couples, particularly women, face stigmatization, discrimination, and divorce. A variety of pathologies are suspected in infertility, which includes endocrine dysfunction, genetic abnormalities, infection and diseases, and autoimmune disorders [7, 8]. These pathologies are triggered by environmental factors, including toxic substance exposure as well as lifestyles, such as delayed marriage, nutrition, obesity, stress, smoking, drug use, and alcohol consumption [9]. An in-depth understanding of these mentioned causes is necessary for the prevention and effective treatment of infertility [7]. The genetic causes, in particular, need more attention and understanding because it

* Correspondence: yahaya.tajudeen@fubk.edu.ng; yahayatajudeen@gmail.com
[1]Department of Biolog, Federal University Birnin-Kebbi, PMB 1157, Birnin-Kebbi, Nigeria
Full list of author information is available at the end of the article accounts for 15–30% of male infertility alone [10, 11]. Fortunately, in the last few decades, technological innovations in biological studies have made possible more insightful findings on the genetic etiology of infertility that may lead to better treatment. This review, therefore, provides an update on genetics and pathophysiology of syndromic and nonsyndromic infertility.

## Main text

### Database searching and search strategy

To identify relevant papers on the topic, academic databases such as PubMed, Google Scholar, Uniport, GeneCards, Genetics Home Reference (GHR), and National Center for Biotechnology Information (NCBI) were searched. Key search words used include 'infertility', 'male infertility', 'female infertility', 'etiology of infertility', and 'causes of infertility'. Others are 'genetic etiology of infertility', 'gene mutations predisposing to infertility', 'syndromic and nonsyndromic infertility', and 'gene mutations causing infertility'. Each database was searched independently, after which the articles retrieved were pooled together and double citations removed.

### Inclusion and exclusion criteria

Articles were included if they are available in the English language, focused on infertility, genetic etiology of infertility, and pathophysiology of infertility. Studies published before the year 2000 were excluded, except sometimes in which the information was vital. This was done to ensure up-to-date information.

A total of 133 articles were identified from all the databases, of which 120 were retained after removing duplicates (Fig. 1). Of the 120 articles retained, 110 passed the relevance test for eligibility. From the eligibility test, 101 articles fit the study objectives and were reviewed and included in this study.

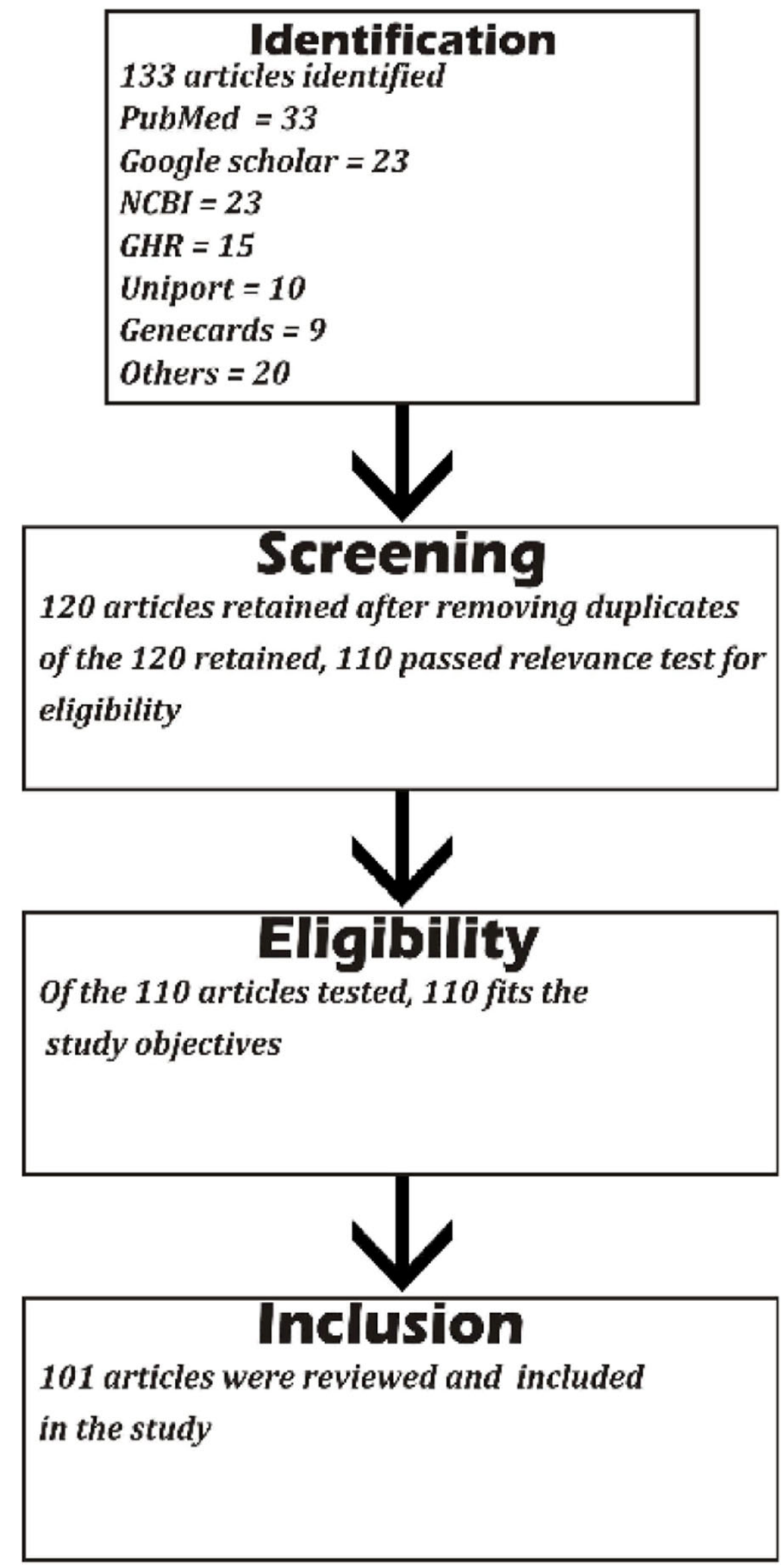

**Fig. 1** Article selection flow diagram

### Genes predisposing to syndromic and nonsyndromic infertility

The searches identified several gene mutations linked with infertility and infertility-related disorders and syndromes. However, it is beyond this study to discuss all the genes. As such, 65 genes frequently encountered in our searches and with sufficient information were included in this study. The genes were classified into genes predisposing to syndromic infertility, genes predisposing to nonsyndromic spermatogenic failure and sperm morphology abnormalities, genes predisposing to nonsyndromic sex reversal and pubertal delay, and genes predisposing to nonsyndromic ovarian failure.

#### *Genes predisposing to syndromic infertility*

Twenty-three (23) genes, representing 35% of the total genes collected, were linked with syndromic infertility (Table 1). The most common syndromes associated with infertility identified by this study are polycystic ovary syndrome (PCOS), Swyer syndrome, and Sertoli cell-only syndrome, respectively. Others include the congenital bilateral absence of the vas deferens (CBAVD), Wilm's tumour, fibroid, Kallmann syndrome, Frasier syndrome, Denys-Drash syndrome, and Bordet-Biedl syndrome. Most of the genes cause female infertility with features such as hypogonadotropic hypogonadism, ovarian failure, sex reversal, gonad underdevelopment, puberty delay, and menstrual disorders. Some genes also predispose to male

**Table 1** Genes predisposing to syndromic infertility

| Gene | Locus | Biological functions | Some mutations reported | Pathophysiology/disorders |
|---|---|---|---|---|
| *CFTR* (cystic fibrosis transmembrane conductance regulator) | 7q31.2 | *CFTR* transports chloride ions into and out of cells, controlling the movement of water in tissues, which is necessary for the production of mucus that lubricates and protects the lining of the airways, digestive system, reproductive system, and other organs and tissues [12]. | A 3-bp deletion named *F508del*, 5 T single nucleotide polymorphism (SNP) within intron 8, 7 T SNP within intron 8, and missense mutation named *R117H* within exon 4 were reported [13]. Others are *G542X*, *G551D*, *R553X*, *W1282X*, and *N1303K* [13, 14]. | It causes CBAVD, which causes a disconnection between the epididymis and the ejaculatory duct, leading to obstructive azoospermia [13]. Also causes cystic fibrosis (CF), which is associated with menstrual irregularities, including amenorrhea, irregular cycles, and anovulation [15]. |
| *NR5A1* (nuclear receptor subfamily 5 group A member 1) | 9q33.3 | *NR5A1* produces a transcription factor called the steroidogenic factor 1 (SF1), which helps control the activity of several genes related to the development of ovaries and testes, particularly the production of sex hormones and sexual differentiation [16]. | A missense heterozygous mutation involving *c. 3G → A* transition and two heterozygous frameshift mutations named *c. 666delC* and *c. 390delG* were reported [17]. The following were also reported: *p. Pro311Leu*, *p. Arg191Cys*, *p. Gly121Ser*, *p. Asp238Asn*, and *p. Gly123Ala/p. Pro129Leu* [18], as well as a heterozygous mutation named *c. 195G > A* [19]. | Predisposes to Swyer syndrome, which disrupts sexual differentiation and prevents affected 46, XY male from developing testes and causing them to develop a uterus and fallopian tubes [16]. Because of the lack of development of the gonads, Swyer syndrome is also called 46, XY complete gonadal dysgenesis. In females, mutations in the gene cause several ovarian anomalies, including 46, XX gonadal dysgenesis [16]. |
| *WT1* (Wilms' tumor 1) | 11p13 | The gene is a transcription factor that is expressed in the kidneys, ovaries, and testes [20] and functions in gonadogenesis. Particularly, it plays an active role in ovarian follicle development [21] and spermatogenesis [22]. | A heterozygous point mutation in intron 7 named *+2, T → G* was reported [23]. *R362Q* and *K386R* missense mutations among Chinese population [22].<br>Moreover, Seabra et al. [24] reported *p. Pro130Leu* and *pCys350Arg* missense mutations among Portuguese. Two heterozygous missense mutations named *p. Pro126Ser* in exon1 and *p. Arg370His* in exon7 were also reported among the Chinese women [25]. A variant named *IVS+4C>T* has also been reported [20]. | The *+2, T → G* mutation causes Wilms' tumor, characterized by congenital male genitourinary malformation [23]. *R362Q* and *K386R* missense mutations cause loss of function of the *WT1* protein, resulting in non-obstructive azoospermia [22]. The *p. Pro126Ser* and *p. Arg370His* missense mutations cause premature ovarian follicles (POF) [25]. Some mutations also cause Frasier syndrome and Denys-Drash syndrome, both of which often affect the male kidney and genitalia development [20]. |
| *FMR1* (fragile X mental retardation 1) | Xq27.3 | The *FMR1* encodes a protein called FMRP, which is expressed in the brain, testes, and ovaries. *FMR1* transmits nerve impulses in the brain. In the cell, it transports mRNA from the nucleus to the sites where proteins are assembled, some of which are necessary for the functioning of the nerves, testes, and ovaries [26]. | A region of the gene contains a *CGG* trinucleotide repeat of less than 10 to about 40. However, the *FMR1* mutation has been reported in which the *CGG* was abnormally repeated from 200 to more than 1000 times [26]. | Abnormal *CGG* expansion causes instability in the region, deactivating the gene and making little or no protein, resulting in a condition called fragile X syndrome characterized mainly by mental retardation [26]. *CGG* elongation between 55 and 200 repeats causes POF [27] and fragile X-associated primary ovarian insufficiency (FXPOI) [26]. FXPOI is characterized by irregular menstrual cycles, early menopause, and elevated levels of follicle-stimulating hormone (FSH) [26]. |
| *GALT* (galactose-1-phosphate uridylyltransferase) | 9p13.3 | The gene synthesizes galactose-1-phosphate uridylyltransferase, which converts galactose obtained from food into glucose, the main fuel for all cellular activities. This chemical reaction also produces an active form of galactose known as UDP-galactose, which is used to build galactose-containing proteins and fats, both which are involved in energy production, chemical signaling, cell structure building, and molecule transport [28]. | The SNP called *Gln188Arg* or *Q188R* is prevalent among white Europeans and North Americans [28]. Another SNP called *Ser135Leu* or *S135L* are found mostly among the African descent [28]. The SNP named *Asn314Asp* or *N314D* was also reported [28]. | It represses or stops the activity of the galactose-1-phosphate uridylyltransferase, preventing cells from converting galactose into glucose. Consequently, galactose-1-phosphate and related compounds build up to toxic levels in the body, damaging tissues and organs, and leading to a condition known as galactosemia [28]. Women with galactosemia express hypergonadotropic hypogonadism and secondary amenorrhea [29], as well as ovarian failure [30]. |
| *GDF9* (growth/differentiation factor 9) | 5q31.1 | This gene encodes a transforming growth factor-beta superfamily, which is necessary for ovarian folliculogenesis and somatic cell function [30, 31]. | Several missense mutations reported [32]. | It elevates the levels of serum gonadotropins and reduces estradiol, predisposing to premature ovarian failure 14 (POF14), an ovarian disorder defined as the cessation of ovarian function under the age of 40 years. The condition is characterized by oligomenorrhea or amenorrhea |

**Table 1** Genes predisposing to syndromic infertility *(Continued)*

| Gene | Locus | Biological functions | Some mutations reported | Pathophysiology/disorders |
|---|---|---|---|---|
| | | | | [33]. Altered expression of the gene is also associated with polycystic ovary syndrome (POS) [31]. |
| *MED12* (mediator complex subunit 12) | Xq13.1 | The *MED12* gene codes for a protein called mediator complex subunit 12, which regulates gene activity by linking transcription factors with an enzyme called RNA polymerase II. The *MED12* protein is involved in several chemical signaling pathways that control many cellular activities, such as cell growth, cell movement, and cell differentiation [34]. | Several somatic mutations in the *MED12* gene have been reported [34]. | It causes uterine leiomyomas, which are noncancerous growths also known as uterine fibroids. Uterine leiomyomas are common among adult women and cause pelvic pain, abnormal bleeding, and, in some cases, infertility [34]. *MED12* mutations produce nonfunctional protein, which disrupts normal cell signaling and impairs regulation of cell growth and other cell functions. As a result, certain cells divide uncontrollably, leading to the growth of a tumor [34]. |
| *ANOS1* (anosmin 1)/ *KAL1* | Xp22.31 | The gene encodes a protein called anosmin-1, which is involved in embryonic development. Anosmin-1 is expressed in the brain and involved in the migration of neurons that produce gonadotropin-releasing hormone (GnRH), which controls the production of several hormones that direct sexual development before birth and during puberty, such as the ovaries and testes functions [35]. | Mutations that delete a part or the entire gene, as well as SNPs that alter or change amino acids in anosmin-1, have been reported [35], among which is *c.1267C>T* [36]. | Alters the synthesis or function of anosmin-1 during embryonic development, resulting in the loss of sense of smell and the production of sex hormone, respectively, and the latter interferes with normal sexual development causing absence or delay of puberty [35]. Mutations in the gene also predispose to Kallmann syndrome, a disorder characterized by hypogonadotropic hypogonadism [35]. Males expressing hypogonadotropic hypogonadism often have an unusually small penis (micropenis), undescended testes, and lack of secondary sex characteristic, while females fail to menstruate and develop breast [35]. |
| *LEP* (leptin) | 7q32.1 | The gene codes for leptin, which is a hormone that takes part in body weight regulation [37], metabolism, and puberty [38], as well as cell signaling that regulates sex development hormones [37]. | Complete deletion of the gene has been reported in infertile humans and rats [38]. A SNP named *rs10244329* was also reported [39]. | Causes congenital leptin deficiency, a disorder that causes the absence of leptin, resulting in the loss of signaling that triggers feelings of satiety, leading to excessive hunger and weight gain, reduced production of hormones that direct sexual development, and ultimately ending in hypogonadotropic hypogonadism [37]. |
| *LEPR* (leptin receptor) | 1p31.3 | The gene synthesizes a protein called leptin receptor, which is embedded in many tissues, including the hypothalamus, and helps regulate body weight by providing binding sites for leptin [40]. | At least 18 *LEPR* gene mutations have been reported [40]. | Results in less receptor protein reaching the cell surface, causing a condition called leptin receptor deficiency, which reduces *LEPR* protein binding and signaling activities as well as satiety, resulting in excessive hunger and weight and reduced sex hormones, culminating in hypogonadotropic hypogonadism [40]. |
| *NR0B1* (nuclear receptor subfamily 0 group B member 1)/ *AHC* (adrenal hypoplasia congenital) | Xp21.2 | The *NR0B1* gene codes for a transcription factor called *DAX1*, which is involved in the development and function of several hormone-producing (endocrine) tissues, including the adrenal glands, hypothalamus, pituitary gland, as well as the ovaries and testes [41]. | Complete and partial deletions of the gene have been reported [41]. Abnormally short versions of the *DAX1* protein as well as SNPs have also been reported [41]. | Produces inactive *DAX1* protein, disrupting normal development and function of hormone-producing tissues, particularly the adrenal glands, hypothalamus, pituitary, and gonads, resulting in a condition called X-linked adrenal hypoplasia congenital [41], characterized by male puberty delay [38]. Mutations in this gene also cause Swyer syndrome [41]. |
| *HESX1* (HESX homeobox 1) | 3p14.3 | The *HESX1* gene encodes a transcription factor that regulates the early embryonic development of several body structures, particularly the pituitary | SNPs as well as insertion and deletion mutations have been reported in this gene [42]. | Alters the function of the *HESX1* protein and represses the activity of other genes, disrupting the formation and early development of the |

**Table 1** Genes predisposing to syndromic infertility *(Continued)*

| Gene | Locus | Biological functions | Some mutations reported | Pathophysiology/disorders |
| --- | --- | --- | --- | --- |
| | | gland [42]. | | pituitary gland, optic nerves, and other brain structures, resulting in gonadotropin deficiency and a condition known as septo-optic dysplasia [42]. Septo-optic dysplasia is characterized by hypogonadotropic hypogonadism [38]. |
| *LHβ* (luteinizing hormone beta-subunit) | 19q13.33 | It encodes the beta subunit of luteinizing hormone (LH), which is expressed in the pituitary gland and promotes spermatogenesis and ovulation by stimulating the testes and ovaries to synthesize steroids [43]. | The SNP named *G1052A* has been reported [44]. Six other SNPs were also identified and are *gC356090A*, *gC356113T*, *gA356701G*, *gG355869A*, *gG356330C*, and *gG356606T* [45]. | Causes defective LH, leading to low testosterone and gonadotropins, culminating in pubertal delay, bilaterally small descended testes, and infertility [38]. Also increases susceptibility to PCOS, characterized by pubertal delay [38]. |
| *LHCGR* (luteinizing hormone/ choriogonadotropin receptor) | 2p16.3 | The gene synthesizes the luteinizing hormone/chorionic gonadotropin receptor which is a receptor for luteinizing hormone and chorionic gonadotropin. In males, chorionic gonadotropin stimulates the development of Leydig cells in the testis, which are also stimulated by luteinizing hormone to produce androgens, such as testosterone that controls male sexual development and reproduction. In females, luteinizing hormone triggers ovulation, while chorionic gonadotropins ensure normal progression of pregnancy [46]. | A SNP called *G935A* has been reported [44]. At least 17 other SNPs were reported [46]. | Impairs the development of *LHCGR* protein, preventing chorionic gonadotropin binding, and resulting in the absence, or poorly developed Leydig cells, a condition called Leydig cell hypoplasia, characterized by low testosterone, which interferes with male sexual development before and after birth [46]. Extreme Leydig cell hypoplasia causes 46, XY male to develop female external genitalia and small undescended testes [46]. Mild Leydig cell hypoplasia results in an external genital that is not clearly male or female [46]. Mutations in the gene also cause polycystic ovary syndrome (POS) [46]. |
| *AR* (androgen receptor) | Xq12 | The *AR* gene produces an androgen receptor, which is expressed in many tissues, where it binds to androgen to form an androgen-receptor complex, which in turn binds to DNA and regulates the activity of certain genes involved in male sexual development [47]. | Abnormal elongation of a DNA segment in the *AR* gene known as *CAG*, which is normally repeated between less than 10 and 36, has been reported [47]. Some SNPs, as well as deletions and insertions, were also reported [47]. | Results in the receptors that are unable to bind androgens or DNA, causing androgen insensitivity syndrome (AIS), a condition that causes male sexual dysfunction before birth and at puberty. The condition also causes 46, XY male sex reversal also known as gonadal dysgenesis [47]. Mutations in the gene also cause polycystic ovary syndrome [47]. |
| *SRY* (sex-determining region Y) | Yp11.2 | The gene encodes a transcription factor called the sex-determining region Y protein, which is located on the Y chromosome and regulates genes involved in male sexual activities, directing a fetus to develop testes and preventing uterus and fallopian tube formation [48]. | Absence and rearrangement that wrongly placed the gene on the X chromosome have been reported [48]. | Prevents production of SRY protein or hampers it, resulting in Swyer syndrome, characterized by 46, XY male sex reversal [48]. Sometimes the mutation may misplace the gene on the X chromosome from the father, causing 46, XX female to develop both ovarian and testicular tissues, a condition called ovotesticular disorder [48]. |
| *VDR* (vitamin D receptor) | 12q13.11 | The gene is expressed in male and female reproductive tissues [49] and synthesizes a protein called vitamin D receptor, which forms a complex with an active form of vitamin D, known as calcitriol, and another protein called retinoid X receptor, which then binds to particular regions of DNA, where it regulates the activity of some genes that control several processes, particularly calcium and phosphate absorption [50]. In mice, *VDR* signaling plays a role in folliculogenesis and fertility [51]. | The SNP in exon 9 named *rs731236* was reported by Bagheri et al. [52]. Szczepańsk et al. [53] also reported two SNPs named *rs1544410* and *rs222857*. | Reduces follicle number, resulting in PCOS [51, 52] or endometriosis-associated infertility [53]. *VDR* knockout in female mice disrupts *VDR* signaling and ovarian response to stimulation, causing defective folliculogenesis and infertility [51]. Male mice deficient of *VDR* showed gonadal insufficiency and decreased sperm count and motility as well as histological abnormalities of the testis [49]. |
| *FKBP4* (FKBP prolyl isomerase 4) | 6p21.3 | The gene encodes *FKBP52*, which plays an important role in potentiating | Deletions and two SNPs, known as *rs2968909* and *rs4409904*, were reported | Causes azoospermia [54] as well as implantation failure and recurrent |

**Table 1** Genes predisposing to syndromic infertility *(Continued)*

| Gene | Locus | Biological functions | Some mutations reported | Pathophysiology/disorders |
|---|---|---|---|---|
| | | androgen receptor (*AR*) signaling in the prostate and accessory glands [54]. | [55]. | pregnancy [56]. Also predisposes to PCOS [55]. |
| *DBY* (DEAD-box Y RNA helicase)/*DDX3Y* (DEAD-box helicase 3 Y-linked) | Yq11.21 | The gene resides in the *AZFa* region on the Y chromosome and is expressed in many tissues, but mostly in the spermatogonia of the testis tissue and translated only in the male germline [57]. | Deletion mutations were reported [57]. | Causes severe testicular pathology known as Sertoli cell-only (SCO) syndrome, a condition that disrupts spermatogenesis [57]. |
| *USP9Y* (ubiquitin specific peptidase 9 Y-linked) | Yq11.221 | *USP9Y* resides in the azoospermia factor a (*AZFa*) region of the Y chromosome and encodes an enzyme called ubiquitin-specific peptidase 9, Y-linked, which is necessary for sperm production [58]. | *AZFa* deletions resulting in complete loss of *USP9Y* have been reported [59]. | Predisposes to Sertoli cell-only syndrome, characterized by the absence of germ cells in the seminiferous tubules, leading to azoospermia [59]. Also causes spermatogenic failure Y-linked 2 (SPGFY2), resulting in azoospermia or oligozoospermia [59]. |
| *PLK4* (Polo-like kinase 4) | 4q28.1 | *PLK4* protein resides in the centrioles and plays an active role in centriolar duplication that is necessary for normal cell division [60, 61]. | A heterozygous mutation called *p.Ile242Asn* was observed in mice [62]. A heterozygous 13 bp deletion called *c.201_213delGAAACATCCTTCT* was also reported [62]. | Causes mitotic error in mice, resulting in patchy germ cell loss in the testes similar to the human Sertoli cell-only syndrome (SCOS) [62, 63]. |
| *BBS9* (Bardet-Biedl syndrome 9) *PTHB1* (parathyroid hormone responsive-B1) | 7p14 | The specific role of the protein released by this gene has not been determined [64]. | A haplotype named *GAAAG* as well as three SNPs named *rs3884597*, *rs6944723*, and *rs11773504* were reported [65]. | Causes Bardet-Biedl syndrome, characterized by many features, including POF [65]. |
| *FSHR* (follicle-stimulating hormone receptor) | 2p16.3 | The gene secretes a receptor for the follicle-stimulating hormone, which functions in the ovary and testis development [66]. | A SNP in exon 7 named *C566T* and involving Ala to Val substitution at residue 189 was reported by Aittomäk [67]. | Predisposes to ovarian dysgenesis 1 (ODG1), characterized by primary amenorrhea, poorly developed streak ovaries, and high serum levels of FSH and LH. May also cause ovarian hyperstimulation syndrome (OHSS), characterized by massive ovarian enlargement as well as multiple serous and hemorrhagic follicular cysts lined by luteinized cells [68]. |

infertility with phenotypic presentations, including hypogonadotropic hypogonadism, sex reversal, puberty delay or absence, gonad underdevelopment, and spermatogenic failure.

### *Genes predisposing to nonsyndromic spermatogenic failure and sperm morphology abnormalities*

Twenty-six (26) genes, representing 40% of the total genes collected, predispose to nonsyndromic spermatogenic failure and sperm morphology abnormalities (Table 2). Most often, mutations in the genes cause meiotic arrest, resulting in acrosome malformation or absence, ultimately ending in sperm head abnormalities such as azoospermia, globozoospermia, oligospermia, and oligozoospermia. In some cases, the meiotic arrest may result in polyploidy spermatozoa, characterized by an enlarged sperm cell head called macrozoospermia. A meiotic arrest may also decrease sperm motility and hyperactivation needed to push spermatozoa through the uterus. Sometimes, mutations in the genes may cause chromatin damage or DNA fragmentation, disrupting spermatogenesis and causing sperm cell structural defects and loss.

### *Genes predisposing to nonsyndromic sex reversal and pubertal delay*

Five (5) of the genes collected, representing 7.69% of the total genes, predispose to sex reversal and puberty delay or absence (Table 3). Most mutations in the genes cause reduced circulating levels of gonadotropins and testosterone, resulting in hypogonadotropic hypogonadism, characterized by the absence or incomplete sexual maturation. Mutations in the genes may also cause complete or partial gonadal dysgenesis, characterized by underdeveloped or presence of both gonads.

### *Genes predisposing to nonsyndromic ovarian failure*

Eleven (11) of the genes collected, representing 16.92% of the total genes, predispose to nonsyndromic ovarian failure (Table 4). Some mutations in the genes may reduce the sensitivity of fully grown immature oocytes to progestin hormone, resulting in a reduced number of

**Table 2** Genes predisposing to nonsyndromic spermatogenic failure and sperm morphological abnormalities

| | | | | |
|---|---|---|---|---|
| *SPATA16* (spermatogenesis-associated 16, also known as NYD-SP12) | 3q26.32 | The gene is expressed mainly in the Golgi apparatus of the cells of testis [69] and actively involved in the formation of sperm acrosome, which plays a role in spermatogenesis and fusion of sperms and eggs [70]. | A homozygous SNP in exon 4 named *c.848G → A* was reported [17]. | Causes acrosome malformation which can be absent in severe cases, resulting in sperm head abnormality characterized by round-headed sperms known as globozoospermia [71]. Also predisposes spermatogenic failure 6 (SPGF6), an infertility disorder caused by spermatogenesis defects [70]. |
| *AURKC* (aurora kinase C) | 19q13.43 | The *AURKC* codes for a protein called aurora kinase, which helps dividing cells separate from each other and ensures the accurate distribution of genetic materials (chromosomes). Aurora kinase C is most abundant in male testes, where it regulates the division of sperm cells, ensuring that every new sperm cell divides accurately and contains one copy of each chromosome [72]. | A homozygous deletion called *c. 144delC* and frequently found among North African descent was reported [72]. | Produces a nonfunctional aurora kinase C or a protein that breaks down quickly, preventing sperm cell division. Consequently, the sperm cells carry extra chromosomes (polyploidy spermatozoa), usually four copies of each instead of the usual one. The increase in chromosome number enlarges the sperm cell head and leads to the presence of multiple tails (flagella), a condition called macrozoospermia. The additional genetic materials may prevent any of the sperm cells from fusing with an egg or may result in miscarriage [72]. |
| *CATSPER* (cation channel sperm associated 1) | 11q13.1 | The gene encodes a protein localized in the tail of sperm cells and transport calcium cations into the cells for normal sperm motility and a type of sperm cell motility called hyperactivation, which is a rigorous movement necessary to push the sperm cells through the cell membrane of the egg cell during fertilization [73]. | Two insertion mutations named *c.539-540insT* and *c.948-949insATGGC*, leading to frameshifts and premature stop codons known as *p.Lys180LysfsX8* and *p.Asp317MetfsX18*, have been reported [74]. | Alters or malfunctions *CATSPER1* protein or produces a protein that is degraded quickly by the cell. This impairs calcium entry into the sperm cell, decreasing the motility and preventing hyperactivation, ultimately resulting in *CATSPER1*-related non-syndromic male infertility. Affected men may also produce a smaller than the usual number of sperm cells or sperm cells that are abnormally shaped [73]. |
| *MTHFR* (methylenetetrahydrofolatereductase) | 1p36.22 | The *MTHFR* gene synthesizes an enzyme called methylenetetrahydrofolate reductase, which converts a form of folate called 5, 10-methylenetetrahydrofolate to an another form called 5-methyltetrahydrofolate. The latter is the primary form of folate in the blood, where it helps converts the amino acid homocysteine to another amino acid called methionine. The body uses methionine to make proteins and other important compound as well as vital in DNA methylation and spermatogenesis [75, 76]. | A SNP named *677C/T* and involving the substitution of an alanine for a valine is the most common in infertile men with *MTHFR* deficiency [13]. The second mutation involved an A to C transition at nucleotide 1298 (*A1298C*), resulting in glutamate to alanine substitution in the *MTHFR* protein [77]. | Loss of *MTHFR* decreases the activity of its enzyme, disrupting folic acid metabolism, resulting in DNA hypo-methylation, ultimately ending in the absence of germinal cells and spermatogenesis arrest [13, 75, 78]. |
| *SYCP3* (synaptonemal complex protein 3) | 12q23.2 | *SYCP3* is embedded in the testis and encoded an essential structural component of the synaptonemal complex, which is involved in synapsis, recombination, and segregation of meiotic chromosomes [79]. | A heterozygous deletion called *643delA*, and a heterozygous genetic change known as *T657C* was identified among Iranian women with recurrent pregnancy losses [17]. | Causes early meiotic arrest, disrupting the spermatogenic process in males [17], resulting in spermatogenic failure 4 (SPGF4), a disorder characterized by azoospermia. In females, early meiotic arrest |

**Table 2** Genes predisposing to nonsyndromic spermatogenic failure and sperm morphological abnormalities *(Continued)*

| | | | | |
|---|---|---|---|---|
| | | The gene ensures centromere pairing during meiosis in male germ cells, thus important for normal spermatogenesis [80]. | | causes recurrent pregnancy loss [79]. |
| *HSF2* (heat-shock transcription factor 2) | 6q22.31 | *HSF2* is expressed in the testis and encodes heat-shock transcription factor 2, which binds specifically to the heat-shock promoter element to activate heat-shock response genes under conditions of heat or other stresses [81]. | Heterozygous missense mutations have been reported [82]. | *HSF2*-null male mice showed embryonic lethality, neuronal defects, and reduced spermatogenesis that relates to meiotic arrest, increased sperm apoptosis, and seminiferous tubule dysgenesis [83]. Male humans showed azoospermia [82]. |
| *SYCP2* (synaptonemal complex protein 2) | 20q13.33 | The gene codes for a major component of the synaptonemal complex, which is required for normal meiotic chromosome synapsis during oocyte and spermatocyte development and for normal male and female fertility [84]. | Heterozygous frameshift mutation and deletion have been reported in the gene [85]. | Alters synaptonemal complex, disrupting spermatogenesis and resulting in cryptozoospermia and azoospermia [85, 86]. |
| *A-MYB/MYBL1* (myeloblastosis oncogene-like 1) | 8q13.1 | *MYBL1* protein is a male-specific master regulator of meiotic genes that are involved in multiple processes in spermatocytes, particularly processes involved in cell cycle progression through pachynema [87]. | A variant named *repro9* involving a C to A transversion at nucleotide 893 of the *MYBL1* mRNA was reported [87]. *MYBL1*−/−, showing meiotic arrest similar to *repro9*, has also been reported [87]. | Causes meiotic arrest in spermatocytes, characterized by defects in autosome synapsis in pachynema, unsynapsed sex chromosomes, incomplete double-strand break repair on synapsed pachytene chromosomes and a lack of crossing over [87]. |
| *TEX11* (testis expressed 11) | Xq13.1 | *TEX11* protein is required for spermatogenesis; particularly certain levels of the protein are required for meiotic progression. The protein is also necessary for normal genome-wide meiotic recombination rates in both sexes [88]. | Frameshift mutations were observed, so also missense mutations, particularly a missense mutation tagged *V748A* was observed among transgenic mice [88]. | Causes meiotic arrest in male mice, resulting in spermatogenic failure, X-linked, 2 (SPGFX2), a disorder characterized by mixed testicular atrophy and azoospermia [89]. Among humans, meiotic arrest leads to non-obstructive azoospermia [88]. |
| *KIT* (v-kit Hardy-Zuckerman 4 feline sarcoma viral oncogene homolog) | 4q12 | The gene is embedded in the reproductive cells and encoded receptor tyrosine kinases, which is involved in signal transduction. The protein takes part in phosphorylation that activates a series of proteins in multiple signaling pathways, which are necessary for normal cell growth, proliferation, survival, and movement in the reproductive cells and certain other cell types [90]. | A SNP in which Asp-816 is replaced with a Val or His residue at exon 17 has been reported [91]. | Causes seminomas and testicular carcinoma [91]. |
| *ADGRG2* (adhesion G protein-coupled receptor G2) | Xp22.13 | This gene encodes an epididymis-specific transmembrane protein, which is involved in a signal transduction pathway controlling epididymal function and male fertility. May particularly regulate fluid exchange within the epididymis [92]. | Three protein-truncating hemizygous mutations, named *c.1545dupT* (*p.Glu516Ter*), *c.2845delT* (*p.Cys949AlafsTer81*), and *c.2002_2006delinsAGA* (*p.Leu668ArgfsTer21*), have been reported [93]. | Predisposes to congenital bilateral aplasia of the vas deferens, X-linked (CBAVDX), a disease characterized by bilateral absence of vas deferens and obstructive azoospermia [92]. |
| *FKBP6* (FKBP prolylisomerase 6) | 7q11.23 | Encodes a protein that functions in immunoregulation, homologous chromosome pairing in meiosis during spermatogenesis and cellular | Deletion in the exon 8 of the gene has been reported [95]. | Causes spermatogenic failure, resulting in azoospermia or severe oligozoospermia [96]. |

**Table 2** Genes predisposing to nonsyndromic spermatogenic failure and sperm morphological abnormalities *(Continued)*

| | | | | |
|---|---|---|---|---|
| | | processes involving protein folding and trafficking [94]. | | |
| *PRM1* (protamine 1) | 16p13.13 | It encodes a protein called protamine 1, which replaces histone during developmental stages of elongating spermatids and compact sperm DNA into a highly condensed, stable, and inactive complex to ensure that quality spermatozoa are produced and as well protect spermatozoa from the degrading effects of free radicals [97]. | A *c.102G>T* transversion that results in the SNP named *p.Arg34Ser*, a missense mutation named *c.119G>A* (*p.Cys40Tyr*), a heterozygous mutation named *c.-107G>C*, and a variant named *c.*51G>C* have been reported [98]. | Increases sperm DNA fragmentation, resulting in oligozoospermia [98]. |
| *PRM2* (protamine 2) | 16p13.13 | *PRM2* secretes protamine 2, which replaces histone during spermatid development, condensing chromatin, and compacting the DNA to ensure production of quality spermatozoa and prevent degradation by free radicals [97]. | A SNP, known as *-190C > A* (*rs2301365*), was identified in both *PRM1* and 2 [99]. | Causes chromatin damage and DNA breaks, resulting in sperm structural defects, reduced motility, and defective spermatogenesis due to haploinsufficiency [100]. |
| *TNP1* (nuclear transition protein 1) | 2q35 | The gene encodes nuclear proteins, which replace nuclear histones and in turn substituted by protamine 1 and 2 during spermatogenesis [101]. | A SNP named *g.IVS1+75T>C* was reported by Heidan et al [102]. A deletion of 15 nucleotides in the 5′-promoter region of the gene was also reported [101]. | Disrupts the highly condensed structure of the sperm nuclear chromatin, resulting in abnormal spermatogenesis [102]. Also causes varicocele, due to the failure of ipsilateral testicular growth and development [102]. |
| *TNP2* (nuclear transition protein 2) | 16p13.13 | *TNP2* participates in the removal of the nucleohistones and in the initial condensation of the spermatid nucleus, thus contributes to the dense packing of spermatid chromatin during spermatogenesis [103]. | A variant named *G1272C* was reported [98]. | *TNP2−/−* in mice affects sperm chromatin structure, causing sperm head abnormalities, acrosome abnormalities in which the acrosomes do not attach to the nuclear envelope, and reduced sperm motility, resulting in tetrazoospermia [103]. |
| *DAZ1* (deleted in azoospermia 1) | Yq11.223 | The gene encodes azoospermia protein 1, which is necessary for spermatogenesis. It binds to the 3′-UTR of mRNAs, regulating their translation, and promoting germ cell progression to meiosis and the formation of haploid germ cells [104]. | *DAZ1* deletions were reported [105]. | Causes Y chromosome infertility known as spermatogenic failure Y-linked 2 (SPGFY2), a disorder resulting in azoospermia or oligozoospermia [104]. Also causes sperm structural abnormalities and reduced motility [106]. |
| *XRCC2* (X-ray repair cross complementing 2) | 7q36.1 | *XRCC2* protein was shown in mice to be required for genetic stability, embryonic neurogenesis and viability [107]. | A SNP in the gene involving *c.41T>C* substitution was reported [108]. | Causes meiotic arrest, resulting in azoospermia [108]. |
| *CCDC62* (coiled-coil domain containing 62) | 12q24.31 | Encodes a nuclear receptor co-activator that enhances estrogen receptor transactivation [109]. The gene is expressed in the acrosome of developing spermatids and mature sperms, showing that it is necessary for spermatogenesis [110]. | A nonsense mutation in the exon 6, which results in the formation of a premature stop codon and a truncated protein, was reported by Li et al. [110]. | Causes defective sperm morphology and reduced motility [110]. |
| *EFCAB9* (EF-hand calcium21 binding domain-containing protein 9) | 5q35.1 | Encodes sperm-specific EF-band domain protein, which is essential for activation of *CATSPER* channel that regulates sperm motility [111]. | *EFCAB9* deletions were reported [111]. | Disrupts *CATSPER* channel signaling, which affects sperm motility [111]. |

**Table 2** Genes predisposing to nonsyndromic spermatogenic failure and sperm morphological abnormalities *(Continued)*

| | | | | |
|---|---|---|---|---|
| *KLHL10* (Kelch-like family member 10) | 17q21 | *KLHL10* encodes a germ cell-specific protein essential for spermatogenesis [63]. | Two missense mutations named *A313T* and *Q216P* were reported [112]. | Impairs homodimerization, resulting in germ cell loss, abnormal spermatids, and severe oligozoospermia [112]. |
| *SEPT12* (septin 12) | 16p13.3 | The gene codes for septin 12, which is expressed exclusively in the testis and involved in spermatogenesis, especially morphogenesis of sperm heads and the elongation of sperm tails [63]. | Two missense mutations named *c.266C>T/p.Thr8 Met* and *c.589G>A/p.Asp197Asn* were reported by Kuo et al. [113]. | Disrupts the structural integrity of sperm by perturbing septin filament formation, causing various sperm abnormalities, including immotility, bent tails, acrosome breakage, round heads, and significant spermatozoa DNA damage, as well as oligoasthenozoosperm and asthenoteratozoospermia [63, 113]. |
| *TAF4B* (TATA-box binding protein associated factor 4b) | 18q11.2 | *TAF4B* encodes a transcriptional coactivator and involved in folliculogenesis, spermatogenesis, and oogenesis [114]. | A nonsense mutation in exon 9 named *p.R611X* was reported [115]. | Causes spermatogenic failure 13 (SPGF13), a disorder resulting in azoospermia or oligozoospermia [114]. |
| *ZMYND15* (zinc finger mynd-containing protein 15) | 17p22.1 | *ZMYND15* codes for a transcription repressor, which in mice is expressed exclusively in the haploid germ cells, particularly during late spermatogenesis [116]. | A mutation in exon 9 of the gene named *p.K507Sfs*3* was reported by Ayhan et al. [115]. | Causes spermatogenic failure 14 (SPGF14), a disorder resulting in azoospermia or oligozoospermia [117]. |
| *NANOS1* (nanos C2HC-type zinc finger 1) | 10q26.11 | This gene encodes a CCHC-type zinc finger protein that is specifically expressed in the germ cells of adult men and regulates the translation by acting as a post-transcriptional repressor [118]. | Two deletion mutations called *p.Pro77_Ser78delinsPro* and *p.Ala173del* have been reported [119]. | Results in spermatogenic failure 12 (SPGF12), an infertility disorder caused by spermatogenesis defects, characterized by decreased sperm motility and concentration, sperm structural defects, non-obstructive azoospermia, oligozoospermia, and oligo-astheno-teratozoospermia [120]. |
| *GALNTL5* (polypeptide N-acetylgalactosaminyltransferase Like 5) | 7q36.1 | *GALNTL5* encodes an inactive protein, which is expressed in the testis and is required during spermatid development in which it participates in protein loading into the acrosomes [121]. | Heterozygous single nucleotide deletion of maternal inheritance was reported [63, 122]. | Decreases glycolytic enzymes, which disrupts protein loading into acrosomes, resulting in asthenozoospermia and poor sperm motility [122]. |

oocytes undergoing meiotic maturation. Mutations in the genes may also cause ovarian dysgenesis, characterized by absence or puberty delay, primary amenorrhea, uterine hypoplasia, and hypogonadotropic hypogonadism. Some mutations prevent the formation of primordial follicles, resulting in reduced oocyte numbers after birth.

In summary, 23 genes, representing 35%, were linked with syndromic infertility, while 42 genes, accounting for 65% cause nonsyndromic infertility. Of the 42 nonsyndromic genes, 26 predispose to spermatogenic failure and sperm morphology abnormalities, 11 cause ovarian failures, and 5 cause sex reversal and puberty delay. Overall, 31 genes (48%) predispose to male infertility, 15 genes (23%) cause female infertility, and 19 genes (29%) predispose to both. The common features of male infertility were spermatogenic failure and sperm morphology abnormalities, while ovarian failure has been the most frequently reported among infertile females. This analysis infers that male genetic infertility was more prevalent than female, with spermatogenic failure and sperm morphology abnormalities being most prevalent.

## Genetic testing for infertility disorders

Knowing the exact cause of infertility allows for better diagnostic decisions and enables enhanced counseling for parents with regard to risks to their children. For this reason, when there is a means, testing of embryos should be recommended for a family with a history of genetic infertility disorders discussed above. Moreover, every healthy-looking individual is a carrier of between 5 to 8 recessive genetic disorders; so the test should be extended to everyone who has the means [157]. It is specifically recommended for embryos of couples who are recessive for a gene infertility disorder.

**Table 3** Genes predisposing to nonsyndromic sex reversal and pubertal delay

| | | | | |
|---|---|---|---|---|
| *GNRHR* (gonadotropin-releasing hormone receptor) | 4q13.2 | This gene encodes the receptor for type 1 gonadotropin-releasing hormone, a receptor that is expressed on the surface of pituitary gonadotrope cells, lymphocytes, breast, ovary, and prostate. *GNRHR* becomes activated after binding with gonadotropin-releasing hormone, and the complex formed causes the release of gonadotropic luteinizing hormones (LH) and follicle-stimulating hormones (FSH) [123]. | At least 19 different mutations have been identified, including heterozygous mutations named *Gln106Arg/Arg262Gln* and *Arg262Gln/Tyr28Cys* [17] as well as homozygous missense mutation named *g. G7167A; p. Arg139His* [124]. | Causes low levels of circulating gonadotropins and testosterone, resulting in hypogonadotropic hypogonadism 7 (HH7), a disorder characterized by absent or incomplete sexual maturation by the age of 18 years [125]. |
| *PROP1* (PROP paired-like homeobox 1) | 5q35.3 | The gene produces a transcription factor embedded only in the pituitary gland and releases hormones for growth, reproduction, and cell differentiation in the pituitary gland [126]. | At least 25 mutations had been reported, the most common of which deletes two amino acids, written as *301-302delAG* [126]. | Reduces pituitary cell differentiation and prevents the release of hormones from the pituitary gland, causing a condition called combined pituitary hormone deficiency, with features like short stature and delayed or absent puberty [126]. |
| *DMRT1* (doublesex- and MAB3-related transcription factor 1) | 9p24.3 | The gene encodes a transcription factor expressed in the testis and involved in male sex determination and differentiation before and after birth by promoting male-specific genes and repressing female-specific genes. May also play a minor role in oogenesis [127]. | Deletions in the gene had been reported [127]. | Predisposes to male-to-female sex reversal in the presence of a normal 46, XY karyotype, referred to as 46, XY sex reversal 4 (SRXY4), characterized by complete or partial gonadal dysgenesis. The mutation may also cause testicular germ cell tumors [127]. |
| *SOX3* (SRY-box transcription factor 3) | Xq27.1 | The gene codes for a transcription factor embedded in the hypothalamus and pituitary gland where it regulates neuronal development and differentiation, and as well promote male sex development [128]. | Copy number variations including two duplications of about 123 kb and 85 kb, a 343 kb deletion immediately upstream of *SOX3*, and a large duplication of approximately 6 Mb that encompasses *SOX3* have been reported [128]. | Causes 46, XX sex reversal 3 (SRXX3), characterized by XX male reversal and a complex phenotype that includes scrotal hypoplasia, microcephaly, developmental delay, and growth retardation [128]. Also causes 46, XX testicular disorder of sex development [128]. |
| *RSPO1* (R-spondin 1) | 1p34.3 | Produces a protein that is essential in ovary determination through regulation of Wnt signaling [129]. | *c.286+1G>A* [130]. | Causes oocytes depletion and masculinized ovaries, resulting in XX true hermaphroditism, also known as an ovotesticular disorder of sexual development, a disorder of gonadal development characterized by the presence of both ovarian and testicular tissue in 46, XX individuals [130]. |

The conventional method used in genetic testing of embryos is the whole sequence amplification. After fertilization, the embryo undergoes mitotic divisions for 5 to 7 days, ending with the development of the blastocyst stage. A biopsy of some blastocysts is done, after which a whole genome amplification of the cells is conducted, usually using polymerase chain reaction [157, 158]. This technique is laborious, time-consuming, and expensive, so recently, a new technique known as the next-generation sequencing is being used for testing genetic disorders in infertile couples and embryos [159]. The protocol is based on an enlarged panel of disease-associated genes (approximately 5000 genes). The large panel of marker genes allows the identifications of a large number of target and non-target genes [157]. However, the technique has some limitations too, which is its inability to detect haploidies, polyploidies, and mosaicisms [157].

## Conclusion

Several studies reviewed showed that certain genes embedded in the hypothalamus, pituitary gland, gonads, and gonadal outflow regulate fertility in both males and females. However, mutational inactivation of these genes may cause syndromic or nonsyndromic infertility. The common features of male infertility include spermatogenic failure, resulting in azoospermia, oligospermia, and chromosome structural abnormalities. Most females express ovarian failure, resulting in menstrual dysfunction

**Table 4** Genes predisposing to nonsyndromic ovarian failure

| | | | | |
|---|---|---|---|---|
| *BMP15* (bone morphogenetic protein 15)/*GDF9B* (growth/ differentiation factor 9B) | Xp11.22 | The gene encodes a member of transforming growth factor-beta superfamily, which plays a role in oocyte maturation and follicular development, through activation of granulosa cells [131]. | Several missense mutations reported [131]. | Causes ovarian dysgenesis 2 (ODG2), a disorder characterized by lack of spontaneous pubertal development, primary amenorrhea, uterine hypoplasia, and hypergonadotropic hypogonadism as a result of streak gonads [132]. May also cause premature ovarian failure 4 (POF4), a disorder in which the ovarian function stops before the age of 40 years and is characterized by oligomenorrhea or amenorrhea, in the presence of elevated levels of serum gonadotropins and low estradiol [131, 132]. |
| *FIGLA* (factor in germline alpha) | 2p13.3 | *FIGLA* encodes a germ cell-specific basic helix-loop-helix transcription factor that regulates the expression of the zona pellucida- and oocyte-specific genes, particularly genes involved in folliculogenesis [30, 133]. | Missense mutations and deletions that resulted in a frameshift had been reported [134]. | *FIGLA* knockout in female mice prevents formation of primordial follicles, and oocyte numbers drop rapidly after birth [135]. May also cause haploinsufficiency, predisposing to premature ovarian failure 6 (POF6), an ovarian disorder defined as the cessation of ovarian function under the age of 40 years and is characterized by oligomenorrhea or amenorrhea, in the presence of elevated levels of serum gonadotropins and low estradiol [134, 136]. |
| *NOBOX* (newborn ovary homeobox) | 7q35 | *NOBOX* encodes a transcriptional regulator with a homeobox motif and is important for early folliculogenesis [137]. | Missense mutations in the homeobox domain were observed in infertile Caucasian or African descent [138]. | Predisposes to POF [139]. |
| *SALL4* (SAL-like 4) | 20q13.2 | The gene is expressed in the testis and oocytes and secretes putative zinc finger transcription factor that plays a role in the pluripotency of oocytes and maintenance of undifferentiated spermatogonia [38, 140]. | Deletions [140]. | Predisposes to nonsyndromic POF [140, 141]. |
| *FSHβ* (follicle-stimulating hormone subunit beta) | 11p14.1 | This gene encodes the beta subunit of the follicle-stimulating hormone which in association with luteinizing hormone induces egg and sperm production [38, 142]. | *Tyr76X*, *Cys51Gly*, and *Val61X* were reported [38]. | Causes low FSH and estradiol, and high LH among females, resulting in the absence or incomplete breast development and sterility [143]. Males produce low androgen, leading to low testosterone and azoospermia, but puberty may be normal or absent [38]. |
| *HCGβ* (human chorionic gonadotrophin) | 19q13.3 | *HCGβ* encodes a hormone called HCG which is secreted mainly by the placenta and is important for normal progression of pregnancy by maintaining the production of steroid hormones and other growth factors in the corpus luteum [144]. | SNPs named *CGB5 p.Val56Leu* (*rs72556325*) and *CGB8 p.Pro73Arg* (*rs72556345*) had been reported [145]. | Causes low levels of HCG during the first trimester of pregnancy, resulting in miscarriage and ectopic pregnancy [146]. |
| *SOHLH1* (spermatogenesis and oogenesis-specific basic helix-loop-helix 1) | 9q34.3 | This gene encodes one of the testis-specific transcription factors which are essential for spermatogenesis, oogenesis, and folliculogenesis. The protein is necessary for spermatogonial proliferation and differentiation as well as regulates both male and female germline differentiation [147]. | Alternatively spliced transcript variants encoding different isoforms have been found for this gene [147]. | Causes ovarian dysgenesis 5 (ODG5), a disorder characterized by lack of spontaneous pubertal development, primary amenorrhea, uterine hypoplasia, and hypergonadotropic hypogonadism [147]. May also result in spermatogenic failure 32 (SPGF32), a condition that is characterized by non-obstructive azoospermia [148]. |

**Table 4** Genes predisposing to nonsyndromic ovarian failure *(Continued)*

| | | | | |
|---|---|---|---|---|
| *SOHLH2* (spermatogenesis and oogenesis specific basic helix-loop-helix 2) | 13q13.3 | The *SOHLH2* is expressed specifically in spermatogonia and oocytes and is required for early spermatogonial and oocyte differentiation [149]. *SOHLH2* is a transcription regulator of both male and female germline differentiation and together with *SOHLH1* regulates oocyte growth and differentiation [150]. | At least 11 mutant variants of *SOHLH2* gene have been reported [151]. In particular, two variants, named *rs6563386* and *rs1328626*, were reported by Song et al. [152]. | *SOHLH2* knockout causes defects in spermatogenesis and oogenesis similar to those in *SOHLH1*-null mice [149]. May also predispose to PO [151]. Some mutant variants may increase the risk of non-obstructive azoospermia [152], as well as the small testis and testicular atrophy [153]. |
| *PGRMC1* (progesterone receptor membrane component 1) | Xq24 | The gene codes for progestin receptor membrane component 1, which associates with and transports a wide range of molecules, including steroids, and the gene has been demonstrated in zebrafish to function in oocyte maturation and meiosis resumption [154]. | Mutant alleles named *ecu4*, *f21*, and *sa37360* were reported in zebrafish (ZFIN) [154]. | *PGRMC1* knockout in zebrafish reduces both spawning frequency and the number of embryos produced by females. It also reduces the sensitivity of fully grown immature oocytes to progestin hormone, resulting in a reduced number of oocytes undergoing meiotic maturation [154]. |
| *ESR1* (estrogen receptor 1) | 6q25.1-q25.2 | Estrogen receptor alpha regulates estrogen action in all reproductive tissues. Estrogen signaling mediates leukemia inhibitory factor expression, which is a cytokine critical for blastocyst implantation [155]. | A SNP named *rs9340799* was reported [155]. | Causes estrogen resistance, resulting in absence of pubertal growth and endometriosis-related infertility [155]. |
| *HES1* (Hes family bHLH transcription factor 1) | 3q29 | *Hes* is expressed in the ovary and encodes transcriptional factors necessary for oocyte survival and maturation [156]. | Deletions were reported [156]. | *HES1* knockout reduces notch signaling and elevates apoptosis, decreasing the number, size, and maturation of oocytes [156]. |

and pregnancy loss. Males and females may also express sex reversal, pubertal delay or absence, and genital abnormalities such as micro-penis and absence of the breast. Male genetic infertility was more prevalent than female, with spermatogenic failure and sperm morphology abnormalities being most prevalent. The mechanisms leading to these pathologies are gene-specific, which, if targeted in the affected, may lead to improved treatment. Medical practitioners are advised to target these genes in the affected.

#### Abbreviations

AIS: Androgen insufficiency syndrome; BBS: Bordet-Biedl syndrome; CBAVD: Congenital bilateral absence of the vas deferens; CF: Cystic fibrosis; FSH: Follicle-stimulating hormone; FXPOI: Fragile X-associated primary ovarian insufficiency; GHR: Genetic home reference; HH: Hypogonadotropic hypogonadism; LH: Luteinizing hormone; NCBI: National Center for Biotechnology Information; OHSS: Ovarian hyperstimulation syndrome; ODG: Ovarian dysgenesis; PCOS: Polycystic ovary syndrome; POF: Premature ovarian failure; SCOS: Sertoli cell-only; SNP: Single nucleotide polymorphism; SPGF: Spermatogenic failure

#### Acknowledgments

Not applicable.

#### Authors' contributions

TOY conceptualized and did literature searches, article writing, and correspondence. UUL, HA, YSK, SSR, ZA, and SA did literature searches, sorting, and referencing. All authors proofread and approved the final manuscript.

#### Funding

Not applicable.

#### Availability of data and materials

Not applicable.

#### Ethics approval and consent to participate

Not applicable.

#### Consent for publication

Not applicable.

#### Competing interests

The authors declare that they have no competing interests.

#### Author details

[1]Department of Biolog, Federal University Birnin-Kebbi, PMB 1157, Birnin-Kebbi, Nigeria. [2]Department of Biochemistry and Molecular Biology, Federal University Birnin-Kebbi, Birnin-Kebbi, Nigeria.

## Publisher's Note

Springer Nature remains neutral with regard to jurisdictional claims in published maps and institutional affiliations.